\documentclass[aps,amsmath,showpacs,floatfix]{revtex4} 
\usepackage{graphicx}
\usepackage{bm}
\begin{document}

\title{Solving the boundary value problem for finite Kirchhoff rods}   

\author{Alexandre F. da Fonseca$^1$ and Marcus A. M. de Aguiar$^2$
\footnote{Corresponding author. email:aguiar@ifi.unicamp.br,
phone:+55-19-37885466} }

\affiliation{$^1$ Instituto de F\'{\i}sica, Universidade de S\~ao
Paulo, USP\\ Caixa Postal 66318, 05315-970, S\~ao Paulo, Brazil}

\affiliation{$^2$ Instituto de F\'{\i}sica `Gleb Wataghin',
Universidade  Estadual de Campinas, Unicamp\\ 13083-970, Campinas, SP,
Brazil}

\begin{abstract} 

The Kirchhoff model describes the statics and dynamics of thin rods
within the approximations of the linear elasticity theory.  In this
paper we develop a method, based on a shooting technique, to find
equilibrium configurations of finite rods subjected to boundary
conditions and given load parameters. The method consists in making a
series of small changes on a {\it trial} solution satisfying the
Kirchhoff equations but not necessarily the boundary conditions. By
linearizing the differential equations around the trial solution we
are able to push its end point to the desired position, step by
step. The method is also useful to obtain configurations of rods with
fixed end points but different mechanical parameters, such as tension,
components of the moment or inhomogeneities.

\end{abstract} 
 
  
\maketitle

\section{Introduction}  
  
The study of conformations of slender elastic rods is of substantial 
utility in several applications, ranging from the fields of structural 
mechanics and engineering to biochemistry and biology.  Examples are 
the study of coiling and loop formation of sub-oceanic cables 
\cite{coyne,zajac,sun,vaz}, filamentary structures of biomolecules 
\cite{zimm,yang,shi,tamar,golds} and bacterial fibers 
\cite{golds2,klapper}, the phenomenon of helix hand reversal in 
climbing plants \cite{goriely} and the shape and dynamics of cracking
whips \cite{alain3}. 
 
The Kirchhoff model \cite{kirchhoff} provides a powerful approach to
study the statics and dynamics of elastic thin rods
\cite{tamar,olson}. In this model the rod is described by a set of
nine partial differential equations (the \emph{Kirchhoff equations})
in the time and arclength of the rod. They contain the forces and
torques plus a triad of vectors describing the deformations of the
rod. These equations are the result of Newton's second law for the
linear and angular momentum applied to the thin rod plus a linear
constitutive relationship between moments and strains. The Kirchhoff
model holds true in the approximation of small curvatures of the rod,
as compared to the radius of the local cross section \cite{dill}. An
interesting characteristic of this model, known as \emph{Kirchhoff
kinetic analogy}, is that the equations governing the static problem
are formally equivalent to the Euler equations describing the motion
of spinning tops in a gravity field. The Kirchhoff equations for
equilibrium configurations can, therefore, be written in hamiltonian
form.
 
The Kirchhoff equations have been solved for a number of simple
situations. Shi and Hearst\cite{shi} first obtained analytical
solutions of the equilibrium equations and, recently, Nizzete and
Goriely \cite{nizzete} completed the study by making a classification
of all kinds of equilibrium solutions. Goriely and Tabor
\cite{tabor1,tabor2} developed a method to study the dynamical
stability of these solutions and Fonseca and de Aguiar \cite{fonseca}
applied this method to study the near equilibrium dynamics of
non-homogeneous closed rods in viscous media. Recently, Tobias \emph{et
al.}  \cite{tobias} developed the necessary and sufficient criteria
for elastic stability of equilibrium configurations of closed rods.
 
In many cases of interest, including biological molecules, the
filaments are subject to boundary conditions. Examples are the
problem of multiprotein structures, such as histones and gyrase, about
which long pieces of DNA wrap \cite{tobias2}, and multiprotein
structures, such as the \emph{lac repressor} complex \cite{mahade}.
Despite the many achievements described above, the study of the
boundary value problem (BVP) associated with Kirchhoff filaments is
still a big challenge. While the integration of differential equations
from initial conditions is a relatively simple numerical task, the
difficulties of finding solutions for given boundary conditions are
well known in classical mechanics, electromagnetism and quantum
mechanics. Typical examples are classical trajectories connecting two
given space points in the time $t$, electric potentials that vanish at
given surfaces and eigenvalues of the Laplace operator defined inside
a finite domain (quantum billiards).

Because of the analogy with spinning tops, the case of trajectories of
Hamiltonian systems is of particular interest here. The {\em monodromy
method}, developed by Baranger and Davis \cite{baranger}, was designed
specifically to find periodic solutions of Hamiltonian systems with
$N$ degrees of freedom. Xavier and de Aguiar \cite{xavier} extended
the method to find non-periodic trajectories with any given
combination of $2N$ position and momenta at initial and/or final
times.

A widely used method to solve BVPs is the so called \emph{shooting
method} \cite{keller,Ha}. For a single second-order differential
equation, the method consists in finding the proper 'velocity' at the
initial point so as to reach the desired 'target' at the end point,
similar to the shooting of a projectile. Examples of applications of
this method to the Kirchhoff equations are the search for homoclinic
orbits in reversible systems \cite{spence}, heteroclinic orbits
resembling tendril pervesions \cite{tyler} and the study of localized
buckling modes of thin elastic filaments \cite{champ1,champ2}.

In the case of open rods some specific BVPs were solved recently.
K\'arolyi and Domokos \cite{domokos}, using symbolic dynamics, found
global invariants for BVPs of elastic linkages, as natural
discretization of continuous elastic beams, an old problem solved by
Euler(see reference \cite{domokos}). Gottlieb and Perkins
\cite{perkins} investigated spatially complex forms in a BVP governing
the equilibrium of slender cables subjected to thrust, torsion and
gravity. Also, the criteria of Tobias \emph{et al.} \cite{tobias} was
applied to linear segments subjected to \emph{strong anchoring end
conditions}, where not only the end points but also the tangent vector
at the end points are held fixed. The dependence of DNA tertiary
structure on end conditions was studied in \cite{tobias2}, where
explicit expressions for equilibrium configurations were obtained for
a specific case with symmetric end conditions. 
 
Our aim in this paper is to develop a method to find equilibrium
solutions of finite rods subjected to boundary conditions at their end
points and with given load parameters. We emphasize that this is
different from the approach in \cite{champ1,champ2}, where the authors
use shooting methods to calculate localized buckling modes. These
modes are treated as homoclinic solutions of the Kirchhoff equations,
corresponding to infinite rods that become asymptotically straight in
the infinite. Our objective is to find equilibrium solutions for
{\em finite} rods subjected to boundary conditions at both ends. 

Our method is an adaptation of the monodromy matrix method for
non-periodic trajectories \cite{xavier} to the hamiltonian formulation
of the static Kirchhoff equations. We work with the Kirchhoff
equations directly in Euler angles, instead of using the Cartesian
position and tangent vectors \cite{champ1,champ2}. The difficulty in
working with Euler angles is that the variables that one wants to hold
fixed, the spatial position of the filament end points, are not the
variables appearing in the differential equations describing the
rod. However, the number of differential equations to be solved is
much smaller in these variables. Using a symmetry of the Hamiltonian,
we end up with only two independent equations to solve. 

One of the motivations of this work is its possible biological
applications as, for example, the study of single DNA molecules
manipulated by optical traps \cite{wuite,meiners,vincent}, and the DNA
loops between multiprotein structures (such as the \emph{lac}
repressor-operator complex) \cite{schleif,mahadevan}.

This work is organized as follows. In Sec. II we review the Kirchhoff
equations and, in Sec. III, their hamiltonian formulation.  In Sec. IV
we describe our method for solving the BVP. The monodromy method
enters as part of the solution, and proves to be a very efficient
tool. In Sec. V we give numerical examples, calculating the three
dimensional configuration of rods with different sets of load
parameters and end positions. We also discuss the existence of
solutions as function of the load parameters. In Sec. VI, motivated by
the repeated sequences of base-pairs commonly found in DNA molecules,
we consider rods with periodically varying Young modulus. We compare
the configurations of these non-homogeneous rods against their
homogeneous counterparts, fixing the same end points and mechanical
parameters. In Sec. VII we summarize our conclusions.

\section{The Kirchhoff Equations}  
  
The Kirchhoff model describes the dynamics of inextensible thin 
elastic filaments within the approximation of linear elasticity theory 
\cite{dill}. They result from the application of Newton's laws of  
mechanics to a thin rod, and consist of two equations describing the
balance of linear and angular momentum plus a constitutive
relationship of linear elasticity theory, relating moments to
strains. The Kirchhoff model assumes that the filament is thin and
weakly bent (\emph{i.e.} its cross-section radius is much smaller than
its length and its curvature at all points). In this approximation it
is possible to derive a one-dimensional theory where forces and
moments are averaged over the cross-sections perpendicular to the
central axis of the filament.
  
A thin tube can be described by a smooth curve ${\bf x}$ in the 3D
space parametrized by the arclength $s$, and whose position depends on
the time: ${\bf{x}}={\bf{x}}(s,t)$. A \emph{local orthonormal basis},
(or \emph{director basis}) ${\bf{d}}_{i}={\bf{d}}_{i}(s,t)$,
$i=1,2,3$, is defined at each point of the curve, with ${\bf{d}}_{3}$
chosen as the tangent vector, ${\bf{d}}_3 = {\bf{x}}'$. In this paper
we shall use primes to denote differentiation with respect to $s$ and
dots to denote differentiation with respect to time. The two
orthonormal vectors, ${\bf{d}}_{1}$ and ${\bf{d}}_{2}$, lie in the
plane normal to ${\bf{d}}_{3}$, for example along the principal axes
of the cross section of the rod. These vectors are chosen such that
$\{{\bf {d}}_{1},{\bf {d}}_{2},{\bf {d}}_{3}\}$ form a right-handed
orthonormal basis for all values of $s$ and $t$. The space and time
evolution of the director basis along the curve are controlled by
\emph{twist} and \emph{spin equations}  
\begin{equation}  
\label{spinTwistEQ}  
{\bf{d}}_{i}'={\bf{k}}\times {\bf{d}}_{i}\, ,\qquad \quad   
\dot{{\bf{d}}}_{i}={\bm{\omega }} \times {\bf{d}}_{i}\qquad \quad i=1,2,3  
\end{equation}  
which follow from the orthonormality relations ${\bf d}_{i} \cdot {\bf
d}_{j}=\delta _{ij}$. The components of ${\bf k}$ and ${\bm \omega}$
in the director basis are defined as ${\bf k}=\sum _{i=1}^{3}k_{i}{\bf
d}_{i}$ and ${\bm \omega} =\sum _{i=1}^{3}\omega _{i}{\bf d}_{i}$.  $
k_{1} $ and $ k_{2} $ are the components of the curvature and $k_{3}$
is the twist density of the rod. The solution of the twist and spin
equations determines $ {\bf d}_{3}(s,t) $, which can be integrated to
give the space curve $ {\bf x}(s,t) $.
  
Let the material points on the rod be labeled by   
\begin{equation}  
\label{posicao}  
{\bf X}(s,t)={\bf x}(s,t)+{\bf r}(s,t),  
\end{equation}  
where   
\begin{equation}  
\label{posicaor}  
{\bf r}(s,t)=x_{1}\; {\bf d}_{1}(s,t)+x_{2}\; {\bf d}_{2}(s,t)  
\end{equation}  
gives the position of the point on the cross section  
\({\mathcal{S}}(s) \), perpendicular to ${\bf x}'(s)$, with respect 
to the central axis. The total force ${\bf F}={\bf F}(s,t)$ and the
total moment ${\bf M}={\bf M}(s,t)$ (with respect to the axis of the
rod) on the cross section are defined by
\begin{equation}  
\label{forca.na.secao.transversal}  
{\bf F}=\int _{{\mathcal{S}}(s)}{\bf p}_{s}\; dS.  
\end{equation}  
\begin{equation}  
\label{torque.do.disco}  
{\bf M}=\int _{{\mathcal{S}}(s)}{\bf r}\times {\bf p}_{s}\; dS,  
\end{equation}  
where ${\bf p}_{s}$ is the contact force per unit area exerted on the cross  
section \( {\mathcal{S}}(s) \). In terms of the director basis we write   
${\bf F}=\sum ^{3}_{i=1}f_{i}{\bf d}_{i}$ and  
${\bf M}=\sum ^{3}_{i=1}M_{i}{\bf d}_{i}$.  
  
In order to derive a set of equations describing the rod as a 
one-dimensional object, the rod is divided into thin disks of length 
\( {\textrm{d}}s \) and cross section \( {\mathcal{S}}(s) \). To each 
of these disks the conservation laws of linear and angular momentum  
are applied \cite{dill}. The result is 
\begin{equation}  
\label{eq1}  
{\bf F}'+\int _{{\mathcal{S}}(s)}{\bf f}_{ext}\; dS 
= \int _{{\mathcal{S}}(s)}\rho_{0}  \ddot{{\bf X}}\;dS,  
\end{equation}  
\begin{equation}  
\label{eq2}  
{\bf M}'+{\bf x}'\times {\bf F}+\int _{{\mathcal{S}}(s)}{\bf r}\times  
{\bf f}_{ext}\;dS  
=\int _{{\mathcal{S}}(s)}\rho _{0} {\bf r}\times \ddot{{\bf X}}\;dS.  
\end{equation}  
where ${\bf f}_{ext}$ is an external force that will not be considered 
in our calculations (${\bf f}_{ext}=0$ in what follows). 
  
In this article we are interested only in the equilibrium solutions 
and, therefore, we shall drop the derivatives with respect to 
time. Assuming that the rod has a uniform circular cross section of 
area $A$, Eqs. (\ref{eq1}) and (\ref{eq2}) can be simplified to yield 
\begin{equation}  
\label{Kir1}  
{\bf F}' = 0,  
\end{equation}  
\begin{equation}  
\label{Kir2}  
{\bf M}' + {\bf d}_{3}\times {\bf F} = 0 
\end{equation}  
which are a set of six equations for 9 variables: ${\bf F}$, ${\bf M}$
and ${\bf k}$ (from which we determine ${\bf d}_i$).  In order to
close the system of equations we need a \emph{constitutive relation}
relating the local forces and moments (stresses) to the elastic
deformations of the body (strains). In the linear theory of
elasticity, and for a homogeneous elastic material, the stress is
proportional to the deformation. The Young's modulus \( E \) and the
Shear modulus \( \mu \) characterize the elastic properties of the
material. Therefore, it is possible to obtain, for small deformations,
a constitutive relation for the moment. For a isotropic material, in
the director basis, this relation is \cite{dill}:
\begin{equation}  
\label{Kir3}  
{\bf M} = EI\left( k_{1}-k_{1}^{u}\right) {\bf d}_{1} +   
EI\left( k_{2}-k_{2}^{u}\right) {\bf d}_{2} +   
2\mu I\left( k_{3}-k_{3}^{u}\right) {\bf d}_{3},  
\end{equation}  
where $I$ is the principal moment of inertia of the cross section, \(
k_{i} \) are the components of the strain vector and \( k^{u}_{i} \)
are the components of the twist vector in the unstressed
configuration. The case \( k^{u}_{i}=0 \) corresponds to a naturally
straight and untwisted rod. We shall assume \( k^{u}_{i}=0 \).
  
Eqs. (\ref{Kir1}), (\ref{Kir2}) and (\ref{Kir3}) can be further simplified  
by the introduction of scaled variables:   
\begin{equation}  
\label{padrao.de.escala}  
\begin{array}{c}  
s\rightarrow sL,  
\quad {\bf F} \rightarrow {\bf F}\frac{EI}{L^2}, \\ \\  
{\bf M}\rightarrow {\bf M}\frac{EI}{L},   
\quad {\bf k} \rightarrow {\bf k}\frac{1}{L}.  
\end{array}  
\end{equation}  
In the new variables the rod has total length $L=1$. The static
Kirchhoff equations become 
\begin{equation}  
\label{Ki1}  
{\bf F}' = 0,  
\end{equation}  
\begin{equation}  
\label{Ki2}  
{\bf M}'+ {\bf d}_{3}\times {\bf F} = 0,  
\end{equation}  
\begin{equation}  
\label{Ki3}  
{\bf M}=k_{1}{\bf d}_{1}+k_{2}{\bf d}_{2}+\Gamma k_{3}{\bf d}_{3}.  
\end{equation}  
where \( \Gamma =2\mu /E \) is an elastic parameter that does not
affect the equilibrium solutions.

\section{Hamiltonian Formulation} 
 
In order to construct a hamiltonian formulation of the Kirchhoff 
equations we first note that Eqs. (\ref{Ki1})-(\ref{Ki3}) are 
integrable if $E$ and $\mu $ are constant \cite{nizzete}. Eq. (\ref{Ki1}) 
shows that the tension ${\bf F}$ is constant. Let us choose the 
direction of the force as the $z$ direction: 
\begin{equation}  
\label{Force_Constant}  
{\bf F}=F{\bf e}_{Z}.  
\end{equation}  
In analogy to the spinning top, the tension ${\bf F}$ corresponds to 
the gravity field $-m{\bf g}$. Here, $F$ can be considered as an 
external parameter and not as a first integral. Substituting Eq. 
(\ref{Force_Constant}) in Eq. (\ref{Ki2}) and projecting along ${\bf 
e}_{Z}$ we get 
\begin{equation}  
\label{int1}  
{\bf M}' \cdot {\bf e}_{Z} \equiv M'_{Z}=0 
\end{equation}  
which does represent a first integral. By projecting the
Eq. (\ref{Ki2}) along ${\bf d}_{3}$ we obtain another integral, 
$M_3$, since
\begin{equation}  
\label{int2}  
{\bf M}' \cdot {\bf d}_{3} \equiv M'_{3}=0.  
\end{equation}  
Finally, it is also possible to show that the elastic energy per unit 
arclength 
\begin{equation}  
\label{int3}  
H = \frac{1}{2}{\bf M} \cdot {\bf k} + {\bf F} \cdot {\bf d}_{3} 
\end{equation}  
is constant, i.e., $H'=0$. Therefore $H$ is the last integral. 
 
The orthonormal Cartesian basis can be connected to the director 
basis by Euler angles with  
\begin{equation}  
\label{fix}  
{\bf d}_{i}=\sum ^{3}_{j=1} S_{ij} \; {\bf e}_{j}  
\end{equation}  
where  
\begin{equation}  
\label{smat}  
S=\left( \begin{array}{ccc}  
\cos \theta \cos \phi \cos \psi -\sin \phi \sin \psi  &   
\cos \theta \cos \phi \sin \psi +\sin \phi \cos \psi  & 
-\cos \phi \sin \theta \\  
-\cos \theta \sin \phi \cos \psi -\cos \phi \sin \psi  &   
-\cos \theta \sin \phi \sin \psi +\cos \phi \cos \psi  & 
\sin \phi \sin \theta \\  
\sin \theta \cos \psi  & \sin \theta \sin \psi  & \cos \theta   
\end{array}\right)   .
\end{equation}  
The static Kirchhoff equations (\ref{Ki1})-(\ref{Ki3}) can then be
written in terms of \( \theta \), \( \phi \) and \(\psi \).  We get
\begin{equation}  
\label{euler}  
\begin{array}{l}  
\theta'' -(\psi' )^{2}\sin \theta \cos \theta +\Gamma \psi' 
(\phi' +\psi' \cos \theta )\sin \theta =  F\sin \theta \\  
\psi'' \sin \theta +2\psi' \theta' \cos \theta -\Gamma \theta' 
(\phi' +\psi' \cos \theta ) =  0\\  
\psi'' \cos \theta  =  \psi' \theta' \sin \theta -\phi''   
\end{array}  \;.
\end{equation}  
  
These equations can also be derived directly from Eqs. 
(\ref{int1})-(\ref{int3}). In terms of the Euler angles the 
Hamiltonian becomes 
\begin{equation}  
\label{hamilton}  
H=\frac{P^{2}_{\theta}}{2}+\frac{P^{2}_{\phi}}{2\Gamma}+  
\frac{(P_{\psi}-P_{\phi}\cos \theta)^{2}}{2\sin ^{2} \theta}+  
F\cos \theta,  
\end{equation}  
where   
\begin{equation}  
\label{Pteta}  
P_{\theta}=\theta ',  
\end{equation}  
\begin{equation}  
\label{Pphi}  
P_{\phi} \equiv M_{3}=\Gamma (\phi '+\psi '\cos \theta),  
\end{equation}  
\begin{equation}  
\label{Ppsi}  
P_{\psi} \equiv M_{Z}=\psi '\sin ^{2}\theta + P_{\phi}\cos \theta .  
\end{equation}  
Eqs. (\ref{euler}) correspond to Hamilton's equations  
$P'_{\alpha}=-\frac{\partial H} {\partial \alpha '}$ and  
$\alpha '=\frac{\partial H}{\partial P_{\alpha}}$ for $\alpha = \theta
,\psi $   or $\phi $. We see immediately that $P_{\phi}$   
and $P_{\psi}$ are constants and that $\theta$ is the only independent 
variable.  
 
The total elastic energy of the rod can be calculated by the integration  
of the Eq. (\ref{hamilton}): 
 
\begin{equation} 
\label{ET} 
E_{T}=\int ^{1}_{0}H(s)ds, 
\end{equation}
The energy is a function of $P_{\psi}$, $P_{\phi}$ and $F$. It also
depends on the initial conditions $\theta (s=0)\equiv \theta _{0}$ and
$P_{\theta}(s=0)\equiv P_{\theta 0}$.
 
The procedure to construct equilibrium solutions for given constants
$P_{\psi}$ and $P_{\phi}$ and initial condition $(\theta
_{0},P_{\theta 0})$ is as follows: first we solve the equations
$P'_{\theta}=-\frac{\partial H} {\partial \theta '}$ and $\theta
'=\frac{\partial H}{\partial P_{\theta}}$ to obtain $(\theta
(s),P_{\theta}(s))$. Second, using Eq. (\ref{Ppsi}), we obtain $\psi
(s)$. The solutions $\theta (s)$ and $\psi (s)$ are sufficient to
construct the rod by integrating the tangent vector ${\bf d}_{3}$:
\begin{equation}  
\label{xdes}  
{\bf x}(s)  =\int^{s}_{0} {\bf d}_{3}(s')ds' \;.
\end{equation}  
Explicitly,
\begin{equation}  
\label{xx}  
x(s)=\int ^{s}_{0}\sin \theta (s')\cos \psi (s')ds' \; ,  
\end{equation}  
\begin{equation}  
\label{yy}  
y(s)=\int ^{s}_{0}\sin \theta (s')\sin \psi (s')ds' \; ,  
\end{equation}  
\begin{equation}  
\label{zz}  
z(s)=\int ^{s}_{0}\cos \theta (s')ds' \; ,
\end{equation} 
where
\begin{equation}
\label{ppsii}
\psi (s)=\psi _{0}+\int ^{s}_{0}\frac{P_{\psi}-P_{\phi}\cos \theta(s')}
{\sin ^{2}\theta(s')}ds'.
\end{equation}

Substituting equation (\ref{ppsii}) in  (\ref{xx}-(\ref{zz}) and
re-arranging the terms we obtain, in matrix form, 
\begin{equation}  
\label{psimatriz}  
\left( \begin{array}{c}  
x(s) \\ y(s) \\ z(s)  
\end{array}\right)  
= \left( \begin{array}{ccc}
\cos \psi _{0} & \sin \psi _{0} & 0 \\
-\sin \psi _{0} & \cos \psi _{0} & 0 \\
0 & 0 & 1 
\end{array}\right)
 \left( \begin{array}{c}  
x _{0}(s) \\ y_{0}(s) \\ z(s)  
\end{array}\right).  
\end{equation}  
where $x_{0}(s)$ and $y_{0}(s)$ are the equations (\ref{xx}) and
(\ref{yy}) for $\psi _{0}\equiv0$. Therefore, it suffices to find the
solution with $\psi _{0}=0$. The solutions for other values of
$\psi_0$ are simple rotations of this basic solution.

\section{The Linearized Method}  
  
In this section we present our method for finding the configuration of
finite rods subject to boundary conditions in the position of its
initial and final points. In fact, since the Kirchhoff equations are
invariant under space translations, we can always choose the initial
point be the origin. As we saw in the previous section, equilibrium
solutions for the static Kirchhoff equations depend only on two
initial conditions, namely, $\theta _{0}$ and $P_{\theta 0}$. The
third initial condition, $\psi_{0}$ corresponds to a rotation of this
solution around the z-axis. The problem is then that of finding a
solution that starts from the origin and ends at $z_{f}$ and at a
distance $r_f\equiv\sqrt{y_{f}^{2}+x_{f}^2}$ from the z-axis. Our
method is based on a series of small deformations made upon an initial
solution of the Kirchhoff equations which, however, does not have the
desired boundary values for $r_f$ and $z_f$.  We call this initial
solution the {\em trial solution}.  The method consists in pushing the
end point of the trial solution to the desired position, step by step.
The basic idea, which is a type of shooting procedure, is to find a
variation in the initial conditions so as to obtain the desired
variation in the end point.

In order to do so, we shall employ a variation of the Monodromy Method
\cite{baranger,xavier}, originally devised to calculate periodic
solutions of chaotic Hamiltonian systems. As discussed in the
Introduction, working with the Kirchhoff equations in Euler angles
poses an extra difficulty on the already hard problem of satisfying
boundary conditions: the variables to be held fixed, $r_f$ and $z_f$,
are not the ones entering the equations of motion, namely, $\theta$,
$P_{\theta}$ and $\psi$. The advantage is that we can find the
solutions solving only two differential equations.
 
The rod can be obtained from the Euler angles the equations
(\ref{xx}), (\ref{yy}) and (\ref{zz}). The Euler angles, in their
turn, obey the equations
\begin{equation}  
\label{eqteta}  
\theta '=P_{\theta}  
\end{equation}  
\begin{equation}  
\label{eqPteta}  
P'_{\theta}=-\frac{(P_{\psi}-P_{\phi}\cos \theta )P_{\phi}}{\sin \theta }  
+\frac{(P_{\psi}-P_{\phi}\cos \theta )^{2}\cos \theta }{\sin ^{3}\theta }  
+F\sin \theta 
\end{equation}  
and 
\begin{equation}  
\label{psides}  
\psi (s)=  
\int ^{s}_{0}\frac{P_{\psi}-P_{\phi}\cos \theta (s')}  
{\sin ^{2} \theta (s')}ds'.   
\end{equation}  
 
If we integrate Eqs.(\ref{eqteta})-(\ref{psides}) using the initial
condition provided by the trial solution and further integrate
Eqs.(\ref{xx})-(\ref{zz}) with the resulting Euler angles, we get, of
course, the trial rod. Variations in these initial conditions will
produce variations in the rod configuration, and, in particular, in
its end point. In what follows we shall construct an explicit relation
between a small variation in the initial variables $\theta _{0}$ and
$P_{\theta 0}$ and the rod's end point, represented by $r_f$ and
$z_f$. Explicitly, we shall find the matrix $B$ such that
\begin{equation}  
\label{dxyzL}  
\left( \begin{array}{c}  
\delta r_f \\ \delta z_f  
\end{array}\right)  
= B \left( \begin{array}{c}  
\delta \theta _{0} \\ \delta P_{\theta 0}   
\end{array}\right).  
\end{equation}  
 
Once $B$ is obtained (and if it can be inverted) we can work our way
from the trial solution, whose end point is at, say, $r_t$ and $z_t$,
to the desired end point at $r_f$ and $z_f$, provided we do that in a
series of small steps. In each step we use the previous solution as
the trial input, pushing the rod's end point slowly towards its final
destination. 

Using $r_f=\sqrt{x_f^2+y_f^2}$, the components of the matrix $B$ can
be written as: 
\begin{equation}
\label{drdte}
B_{11} = \frac{\partial r_f}{\partial \theta_{0}}=
\frac{x_f}{r_f}\frac{\partial x_f}{\partial \theta_{0}} +
\frac{y_f}{r_f}\frac{\partial y_f}{\partial \theta_{0}} 
\end{equation}
\begin{equation}
\label{drdPte}
B_{12} = \frac{\partial r_f}{\partial P_{\theta 0}}=
\frac{x_f}{r_f}\frac{\partial x_f}{\partial P_{\theta 0}} +
\frac{y_f}{r_f}\frac{\partial y_f}{\partial P_{\theta 0}} 
\end{equation}
\begin{equation}
\label{b21}
B_{21} = \frac{\partial z_f}{\partial \theta _{0}} 
\end{equation}
\begin{equation}
\label{b22}
B_{22} = \frac{\partial z_f}{\partial P_{\theta 0}} \;.
\end{equation}

From Eqs. (\ref{xx})-(\ref{zz}) and (\ref{psides}) we find
\begin{equation}  
\label{dxf}  
\delta x_{f}=  
\int ^{1}_{0}\cos \theta (s)\cos \psi (s)\delta \theta (s)ds-  
\int ^{1}_{0}\sin \theta (s)\sin \psi (s)\delta \psi (s)ds,  
\end{equation}  
\begin{equation}  
\label{dyf}  
\delta y_{f}=  
\int ^{1}_{0}\cos \theta (s)\sin \psi (s)\delta \theta (s)ds+  
\int ^{1}_{0}\sin \theta (s)\cos \psi (s)\delta \psi (s)ds , 
\end{equation}  
\begin{equation}  
\label{dzf}  
\delta z_{f}=-  
\int ^{1}_{0}\sin \theta (s)\delta \theta (s)ds ,  
\end{equation}  
and 
\begin{equation}  
\label{dpsi}  
\delta \psi (s)=  
\int ^{s}_{0}A(\theta (s'))\delta \theta (s')ds',  
\end{equation}  
where $A(\theta )$ is given by   
\begin{equation}  
\label{Ateta}  
A(\theta )=\frac{P_{\phi}}{\sin \theta}-  
\frac{2(P_{\psi}-P_{\phi}\cos \theta )\cos \theta }{\sin ^{3}\theta }.  
\end{equation}  
 
Finally, to find the relation between the variations $\delta
\theta(s)$ and  $\delta P_{\theta}(s)$ and their values at the initial
point $s=0$, we  consider small variations of Eqs.(\ref{eqteta}) and
(\ref{eqPteta}) around the trial solution : 
\begin{equation}   
\label{dtedPte}  
\begin{array}{l}  
\delta \theta '=\delta P_{\theta}, \\  
\delta P'_{\theta}=C(\theta )\delta \theta , 
\end{array}  
\end{equation}   
where $C(\theta )$, given by  
\begin{equation}   
\label{Cteta}  
C(\theta )=-P^{2}_{\phi}-  
\frac{(P_{\psi}-P_{\phi}\cos \theta )(P_{\psi}-4P_{\phi}\cos \theta )}  
{\sin ^{2}\theta }-  
\frac{3(P_{\psi}-P_{\phi}\cos \theta )^{2}\cos ^{2}\theta }{\sin ^{4}\theta }  
+F\cos \theta , 
\end{equation}   
is computed at the trial solution. 
 
The solution to these linear equations can be written in matrix form as 
\begin{equation}  
\label{Mteta}  
\left( \begin{array}{c}  
\delta \theta(s) \\ \delta P_{\theta }(s)  
\end{array}\right)  
=\left( \begin{array}{cc}  
M_{11}(s) & M_{12}(s) \\  
M_{21}(s) & M_{22}(s)   
\end{array}\right)  
\left( \begin{array}{c}  
\delta \theta _{0} \\ \delta P_0
\end{array}\right),  
\end{equation} 
where $M$ is the {\em tangent matrix}, satisfying $M(0)=1$. In the 
special case where the trial solution is periodic, $M$ is called the 
{\em monodromy matrix}. 

Writing $\delta \theta (s)$ explicitly as  
\begin{equation}  
\label{dteta}  
\delta \theta (s)=M_{11}(s)\delta \theta _{0}+  
M_{12}(s)\delta P_0,  
\end{equation}  
and using Eqs.(\ref{dxf})-(\ref{dpsi}) we can readily obtain
\begin{equation}
\label{bb}
\begin{array}{l}  
\displaystyle{
\frac{\partial x_f}{\partial \theta_{0}}=
\int ^{1}_{0}\cos \theta (s)\cos \psi (s)M_{11}(s)ds-  
\int ^{1}_{0}\sin \theta (s)\sin \psi (s)  
\int ^{s}_{0}A(\theta (s'))M_{11}(s')ds' ds }\\  
\displaystyle{
\frac{\partial x_f}{\partial P_{\theta 0}}=
\int ^{1}_{0}\cos \theta (s)\cos \psi (s)M_{12}(s)ds-  
\int ^{1}_{0}\sin \theta (s)\sin \psi (s)  
\int ^{s}_{0}A(\theta (s'))M_{12}(s')ds' ds }\\  
\displaystyle{
\frac{\partial y_f}{\partial \theta_{0}}=
\int ^{1}_{0}\cos \theta (s)\sin \psi (s)M_{11}(s)ds+  
\int ^{1}_{0}\sin \theta (s)\cos \psi (s)  
\int ^{s}_{0}A(\theta (s'))M_{11}(s')ds' ds }\\  
\displaystyle{
\frac{\partial y_f}{\partial P_{\theta 0}}=
\int ^{1}_{0}\cos \theta (s)\sin \psi (s)M_{12}(s)ds+  
\int ^{1}_{0}\sin \theta (s)\cos \psi (s)  
\int ^{s}_{0}A(\theta (s'))M_{12}(s')ds' ds} \\  
\displaystyle{
\frac{\partial z_f}{\partial \theta_{0}}=
-\int ^{1}_{0}\sin \theta (s)M_{11}(s)ds }\\  
\displaystyle{
\frac{\partial z_f}{\partial P_{\theta 0}}=
-\int ^{1}_{0}\sin \theta (s)M_{12}(s)ds }\\  
\end{array}  
\end{equation}  
and, therefore, the matrix $B$,

Since we linearized the equations of motion, we have to check if the
new solution, starting from $\theta _{1}=\theta _{0}+\delta \theta
_{0}$ and $P_{\theta 1}=P_0+\delta P_{\theta 0}$ generates a rod with
the chosen final point, within a given precision. If the precision is
not reached, we can use the newly computed solution as a new trial
one, using again Eq. (\ref{dxyzL}), now with ($\delta r_f, \delta
z_f$) corresponding to the distance between the fixed end point and
the end point of the previously computed rod. The process can be
repeated until the desired accuracy is obtained.
  
Finally we note that the elements $M_{ij}(s)$ can be computed by
solving the linear equations (\ref{dtedPte}) with proper initial
conditions. Indeed, setting $\delta \theta _{0}=1$ and $\delta
P_{\theta 0}=0$, Eq. (\ref{Mteta}) gives $M_{11}(s)=\delta \theta (s)$
and $M_{21}(s)=\delta P_{\theta }(s)$.  If, on the other hand, we set
$\delta \theta _{0}=0$ and $\delta P_{\theta 0}=1$ we get
$M_{12}(s)=\delta \theta (s)$ and $M_{22}(s)=\delta P_{\theta }(s)$.
Therefore, $M_{11}(s)$ and $M_{21}(s)$ are solutions of the linearized
Eqs.  (\ref{dtedPte}) with the initial conditions $\delta \theta
_{0}=1$ and $\delta P_{\theta 0}=0$ and $M_{12}(s)$ and $M_{22}(s)$
are the solutions of the same equations with $\delta \theta _{0}=0$
and $\delta P_{\theta 0}=1$.
  
In many cases we might want to push the rod's end-point to a position
${\bf r}_f=(r_f,z_f)$ far from that of the initial trial solution,
${\bf r}_t=(r_t,z_t)$. To do that we can divide the line connecting
${\bf r}_t$ to ${\bf r}_f$ into $N$ small segments and apply the
linearized method $N$ times, moving a small distance at each step. The
number of steps required will depend on the particular configuration
and possibly on the stability of the rod. In all integrations
presented in this paper, we used a fourth order Runge-Kuta method with
fixed step. In all cases the distance between the end-point of the
trial rod and the target position was divided into $10$ segments and
the solution converged to the desired boundary condition with a
precision of $10^{-6}$ in each component $r$ and $z$.

\begin{figure}[ht] 
  \begin{center}
  \includegraphics[height=145mm,width=65mm,clip]{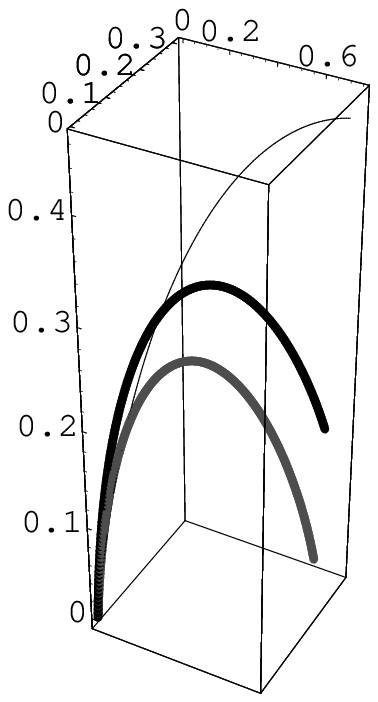}
  \end{center} \caption{Trial (thin line), intermediate (thick line)
  and final (thick gray line) solutions for a rod with desired end
  point at ${\bf r}_f=(0.7,0.0)$. The load parameters are
  $P_{\psi}=1$, $P_{\phi}=1$ and $F=1$, in scaled units.}
\label{fig1}
\end{figure}

Figure \ref{fig1} shows a example of the method. We have chosen the
following load parameters in scaled units (see
Eq. (\ref{padrao.de.escala}): $P_{\psi}=1.0$, $P_{\phi}=1.0$ and
$F=1.0$. The desired end point is $r_f=0.7$ and $z_f=0.0$. We plot the
trial, an intermediate and the converged rods together, in order to
show the process of convergence from the trial to the desired
solution. The trial solution was computed integrating the Kirchhoff
equations starting from $\theta_0=0.5$ and $P_{\theta 0}=1.0$, which
corresponds to a rod whose final point is $r_f\simeq0.8$ and
$z_f\simeq0.455$. The intermediate solution was computed from
$\theta_0\simeq0.216$ and $P_{\theta 0} \simeq 2.257$, which
corresponds to a rod whose final point is $r_f\simeq0.73$ and
$z_f\simeq0.15$.  Finally, the initial conditions obtained for the
converged solution are $\theta _{0}\simeq 0.307$ and $P_{\theta 0}
\simeq 2.510$.

\section{Numerical Examples}  

The particular trial solution used in the previous numerical example
converged smoothly to the chosen final position. In some cases,
however, a given trial solution does not converge to its destination
no matter how many intermediate steps are used to divide the line
between ${\bf r}_t$ and ${\bf r}_f$. As we shall see, this problem has
to do with the existence or not of solutions for a given ${\bf
r}_f$. In scaled units, it is obvious that there are no solutions if
$\sqrt{r_{f}^{2}+z_{f}^{2}}>1$. The restrictions are actually much
stronger than this simple 'length rule', and depend on the values of
$P_{\phi}$, $P_{\psi}$ and $F$. It might also happen that the solution
for a given ${\bf r}_f$ does exist, but that the straight line
connecting ${\bf r}_t$ to ${\bf r}_f$ passes through {\em forbidden}
regions, hindering the convergence. In this section we investigate the
space of possible solutions and give several examples of rods computed
with our method.
 
Each initial condition $\theta_0$ and $P_{\theta 0}$ leads to an end
point ${\bf r}_f$. The easiest way to map all possible final points is
to scan the space of initial conditions. Therefore, for a fixed set of
parameters $P_{\psi}$, $P_{\phi}$ and $F$ we calculate ${\bf r}_f =
{\bf r}_f (\theta_0, P_{\theta 0})$ and plot the resulting figure in
the $(r_f,z_f)$ space. Points outside this region are unreachable by
the rod. Changing the parameters, such as the tension, changes the
region of possible solutions, including end-points that were not
previously present and excluding others.

The results in this section are presented as follows: for each fixed
set of $P_{\psi}$, $P_{\phi}$ and $F$ we show the $(r_f,z_f)$-space of
possible solutions. On the same plot we draw curves of constant
$D=\sqrt{r_{f}^{2}+z_{f}^{2}}$ and, for each $D$ we compute the three
dimensional configuration of a few rods.

In order to compare the rods, we always adjust the value of $\psi_{0}$
such that the rod ends in  $y_f = 0$ plane.  $\psi_{0}$ is determined
by:
\begin{equation}
\label{psi00}
\tan \psi_{0}=\frac{y_{f}^{0}}{x_{f}^{0}} 
\end{equation}
where $x_{f}^{0}$ and $y_{f}^{0}$ are the final values of $x$
and $y$ for the rod calculated with $\psi_{0}=0$. 

In all cases tested our method converged with at least six significant
figures to the previously defined final values of $r_f$ and $z_f$.

\subsection{$P_{\psi}=0, P_{\phi}=1$ and $F=0.1$}

\begin{figure}[ht] 
  \begin{center}
  \includegraphics[height=150mm,width=150mm,clip]{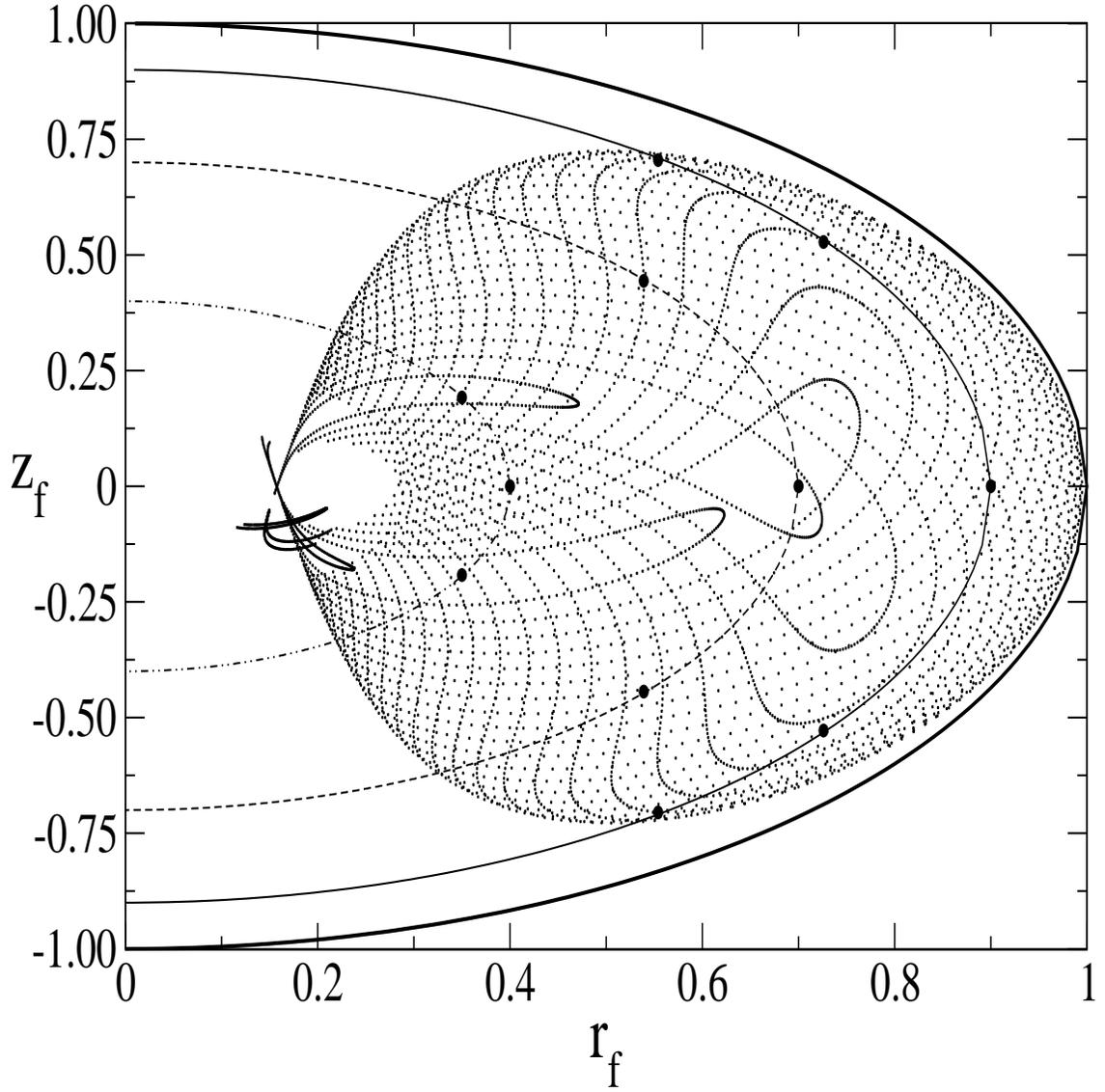}
  \end{center} \caption{Regions of the existence of final points for
  $P_{\psi}=0$, $P_{\phi}=1$, $F=0.1$. The curves $D=1.0$ (full
  thick line), $D=0.9$ (full line), $D=0.7$ (dashed line) and $D=0.4$
  (dotted-dashed line) are also shown. The circles correspond
  to the rods in the Figure \ref{fig3}.The plot was generated by
  varying the initial conditions in the intervals 
  $0<\theta _{0}<\pi $ ($30$ points) and $-5.0<P_{\theta 0}<5.0$
  ($200$ points).}
\label{fig2}
\end{figure}

Figure \ref{fig2} shows the $(r_f,z_f)$ map for this case.  The map
was generated by varying the initial condition in the intervals
$0<\theta _{0}<\pi $ ($30$ points) and $-5.0<P_{\theta 0}<5.0$ ($200$
points). For larger values of $P_{\theta 0}$ the total elastic energy
of the rod increases and the final points tend to concentrate in the
region near the origin (data not shown). The full thick line
represents the curve $D=1.0$, which is the natural limit for the
solutions. But there is a large region inside the $D=1$ line where no
solutions exist. We shall compare it with that of other load
parameters later on. We also show the lines of constant $D$ for
$D=0.9$ (full line), $D=0.7$ (dashed line) and $D=0.4$ (dotted-dashed
line). It is also interesting to note a forbidden region centered
around $z_f\sim0$ and $r_f\sim0.25$. From the sequence of points
crossing each other, it is evident that there are two sets of initial
conditions $(\theta_{0},P_{\theta 0})$ that generate rods with the
same end point. In general they correspond to rods that are above or
below the $z$ axis, and we shall call them the 'up' solution and the
'down' solution respectively.

\begin{figure}[ht] 
  \begin{center}
  \includegraphics[height=95mm,width=50mm,clip]{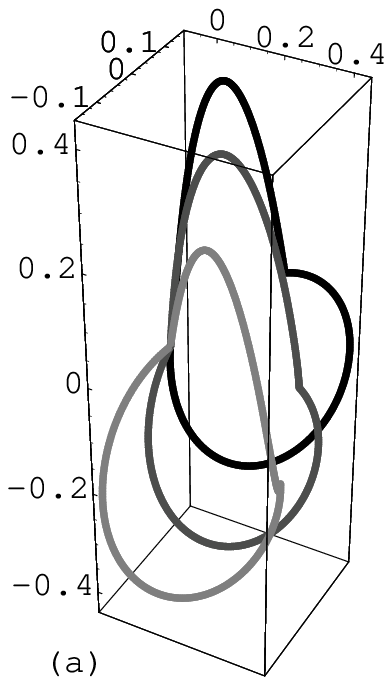}
  \includegraphics[height=95mm,width=50mm,clip]{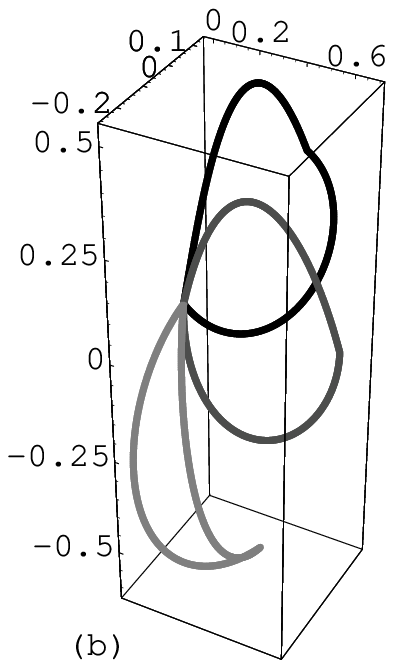}
  \includegraphics[height=95mm,width=50mm,clip]{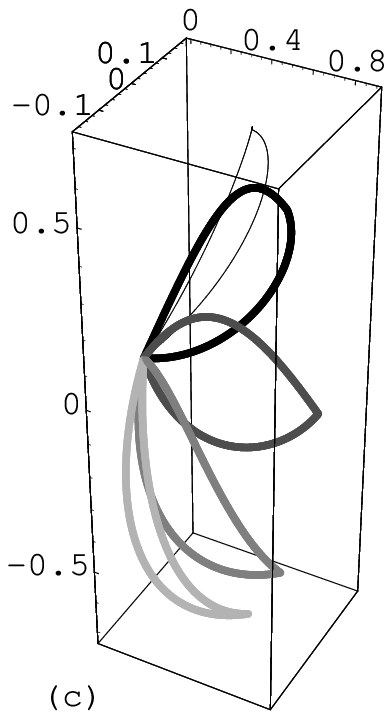}
  \end{center} \caption{Up and down solutions (drawn in the same
  color) for the end points shown by the circles in
  Fig. \ref{fig2}. (a) $D=0.4$; (b) $D=0.7$; and (c) $D=0.9$.} 
\label{fig3} 
\end{figure}

Figure \ref{fig3} shows examples of the rods whose end points are
marked with circles in Fig. \ref{fig2}. We show the up and down
solutions for each final point. Figs. \ref{fig3}a and \ref{fig3}b show
three rods each for $D=0.4$ and $D=0.7$. Fig. \ref{fig3}c shows 5
different rods for $D=0.9$.

Notice that when $P_{\phi}$ or $P_{\psi}$ are zero, the hamiltonian
(\ref{hamilton}), becomes symmetric under the transformation $F
\rightarrow -F$ and $\theta \rightarrow \pi-\theta$. In these cases
the 'up' solution for a given $F$ and $\theta_0$ is identical to the
'down' solution for $-F$ and $\pi-\theta_0$. When both $P_{\phi}$ and
$P_{\psi}$ are non-zero the symmetry disappears.

\subsection{$P_{\psi}=0, P_{\phi}=5$ and $F=1$}

\begin{figure}[ht] 
  \begin{center}
  \includegraphics[height=150mm,width=150mm,clip]{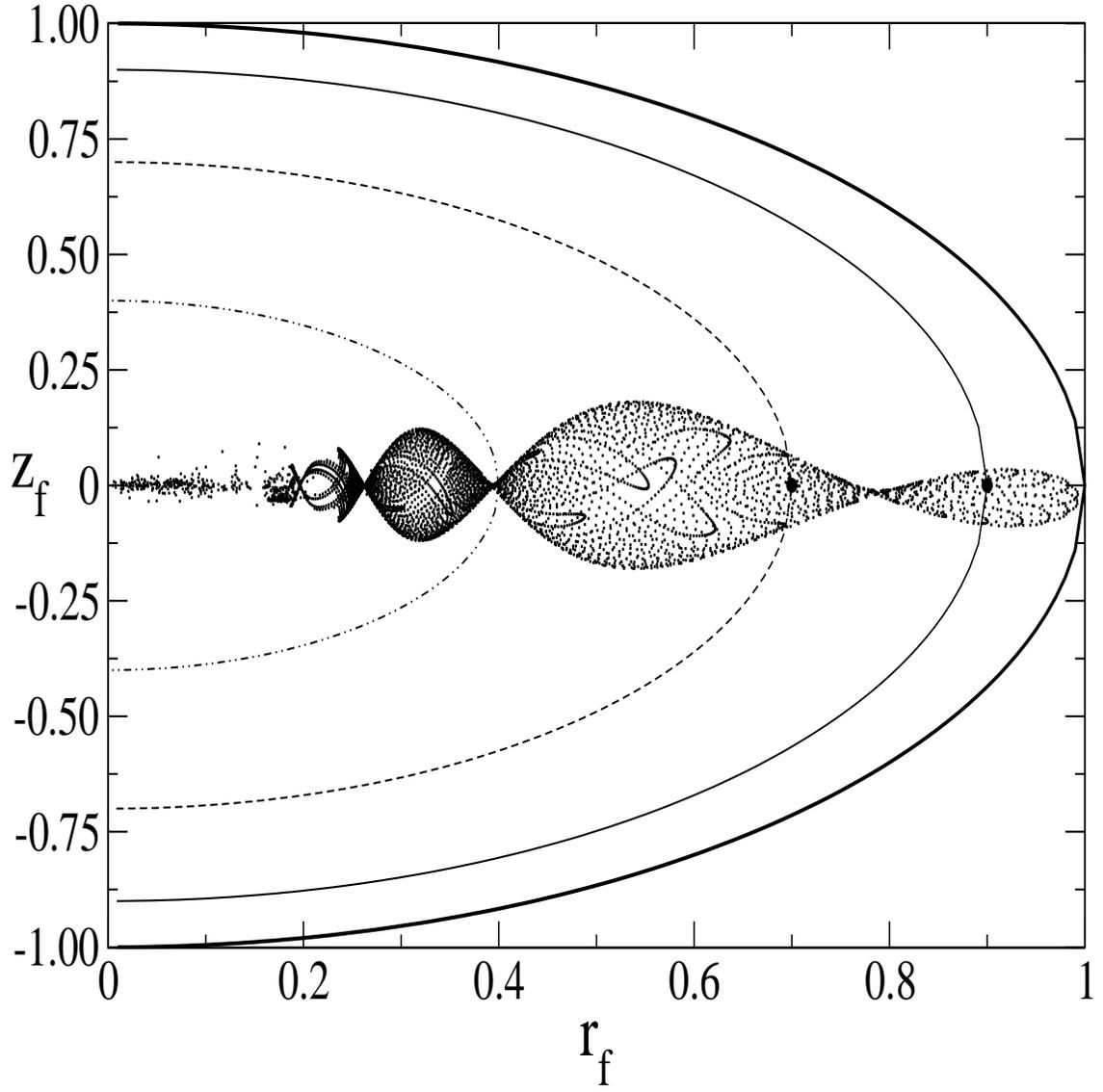}
  \end{center} \caption{Regions of the existence of final points for
  $P_{\psi}=0$, $P_{\phi}=5$, $F=1$. The curves $D=1.0$ (full
  thick line), $D=0.9$ (full line), $D=0.7$ (dashed line) and $D=0.4$
  (dotted-dashed line) are also shown. The circles indicate the 
  rods drawn in Figure \ref{fig5}. The plot was generated by
  varying the initial conditions in the intervals 
  $0<\theta _{0}<\pi $ ($30$ points) and $-20.0<P_{\theta 0}<20.0$
  ($200$ points).}
\label{fig4}
\end{figure}

The $(r_f,z_f)$ map for these parameters is shown in
Fig. \ref{fig4}. It has a curious pattern of thin bulbs centered
around the $z_f=0$ axis, that degenerate for small $r_f$. The only
rods possible in this case are those that return close to the $z=0$
plane.

\begin{figure}[ht] 
  \begin{center}
  \includegraphics[height=95mm,width=60mm,clip]{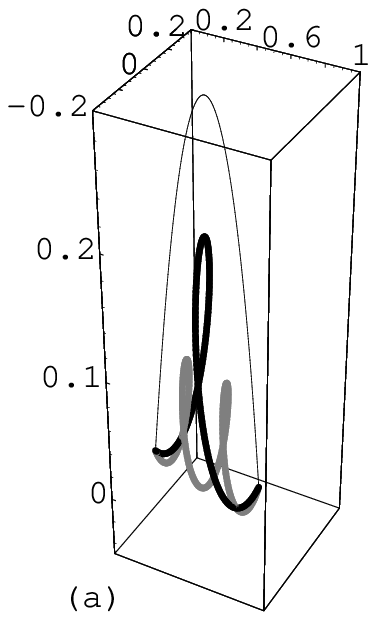}
  \includegraphics[height=95mm,width=60mm,clip]{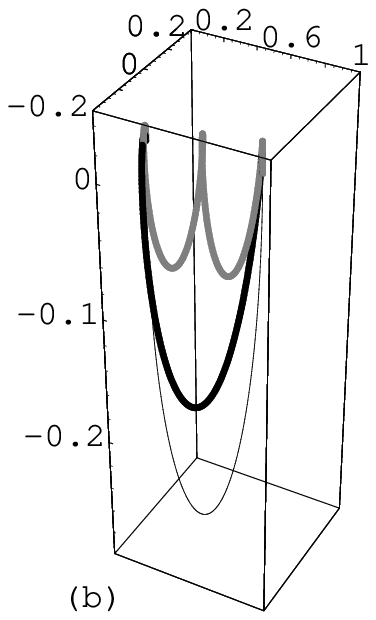}
  \includegraphics[height=95mm,width=60mm,clip]{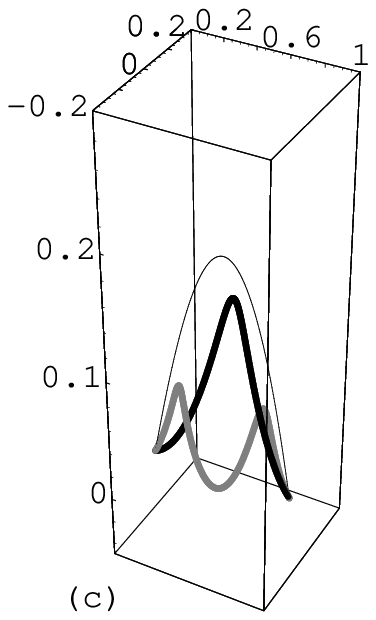}
  \includegraphics[height=95mm,width=60mm,clip]{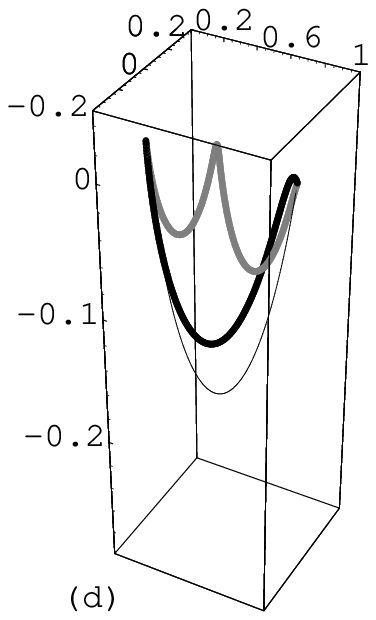}
  \end{center} \caption{Rods for $P_{\psi}=0$ and $F=1$. The lines
  corresponds to $P_{\phi}=1$ (thin black), $P_{\phi}=5$ (thick black)
  and $P_{\phi}=10$ (gray). (a) up solutions for $r_f=0.7$; (b) down 
  solutions for $r_f=0.7$; (c) up solutions for $r_f=0.9$; (d) down
  solutions for $r_f=0.9$.}
\label{fig5} 
\end{figure}

Keeping $P_{\psi}=0$ and $F=1$ but increasing $P_{\phi}$ results in
even thiner bulbs. Figures \ref{fig5} (a) and (b) show the up and down
solutions, respectively, for $r_f=0.7$, $z_f=0$ and $P_{\phi}=1$, $5$
and $10$. Panels (c) and (d) show the up and down solutions for the
same parameters, except for $r_f=0.9$. Notice that large values of
$P_{\phi}$ correspond to horizontal helical rods.

\subsection{$P_{\psi}=P_{\phi}$ and $F=1$}

\begin{figure}[ht] 
  \begin{center}
  \includegraphics[height=150mm,width=150mm,clip]{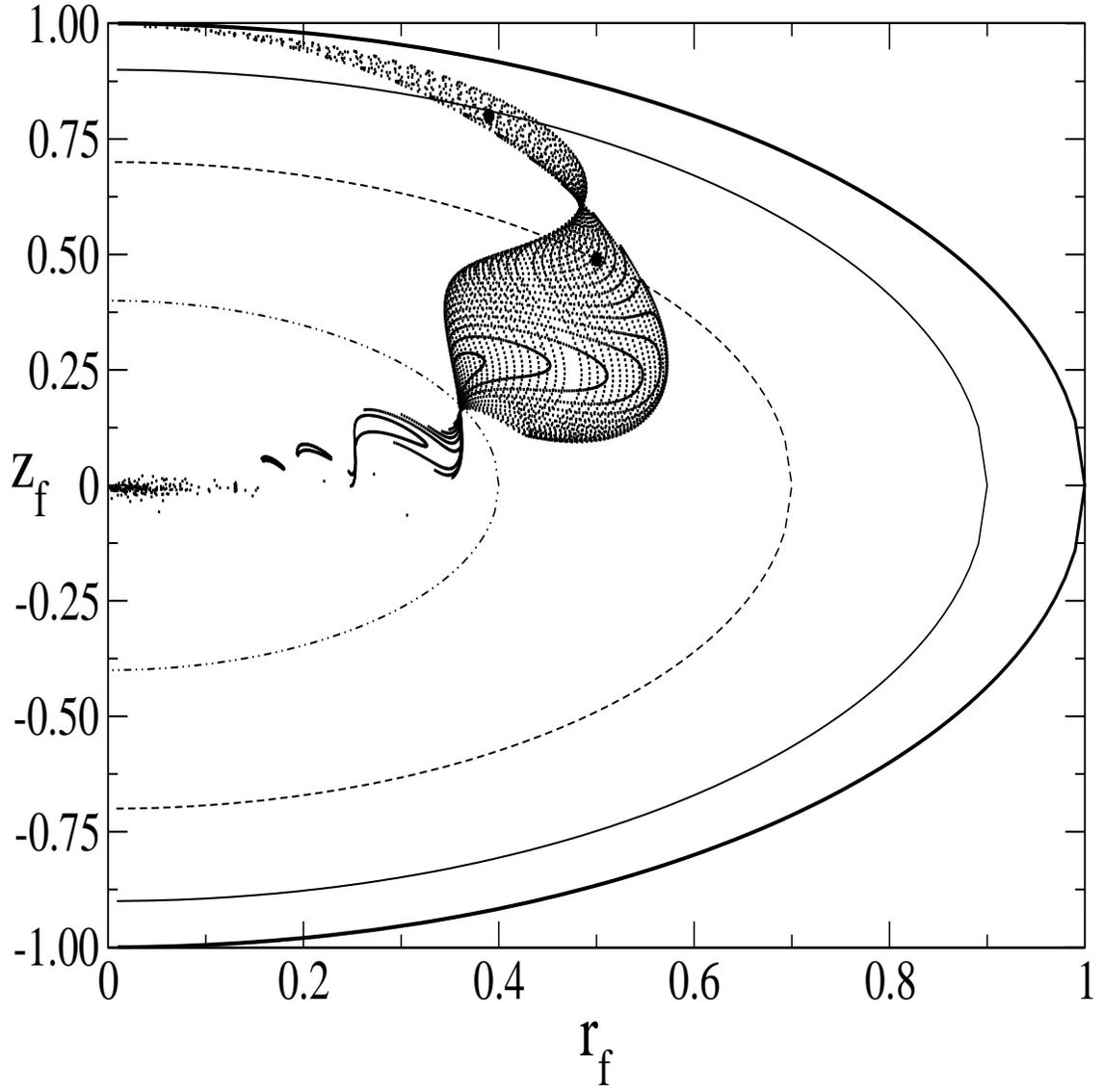}
  \end{center} \caption{Regions of the existence of final points for
  $P_{\psi}=P_{\phi}=5$, $F=1$. The curves $D=1.0$ (full
  thick line), $D=0.9$ (full line), $D=0.7$ (dashed line) and $D=0.4$
  (dotted-dashed line) are also shown. The circles indicate the 
  rods drawn in Figure \ref{fig7}. The plot was generated by
  varying the initial conditions in the intervals 
  $0<\theta _{0}<\pi $ ($30$ points) and $-10.0<P_{\theta 0}<10.0$
  ($200$ points).}
\label{fig6} 
\end{figure}

\begin{figure}[ht] 
  \begin{center}
  \includegraphics[height=100mm,width=60mm,clip]{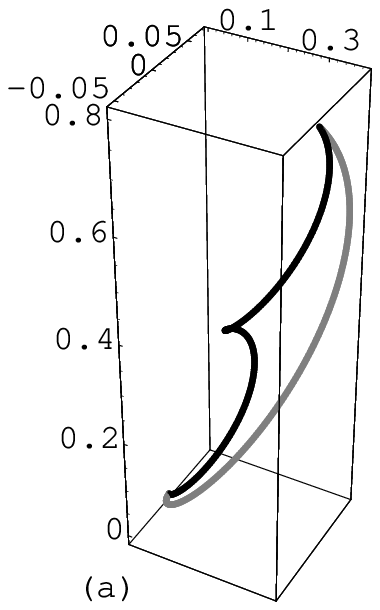}
  \includegraphics[height=100mm,width=60mm,clip]{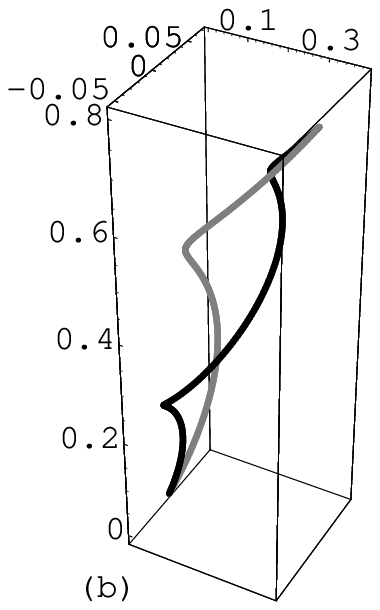}
  \includegraphics[height=100mm,width=60mm,clip]{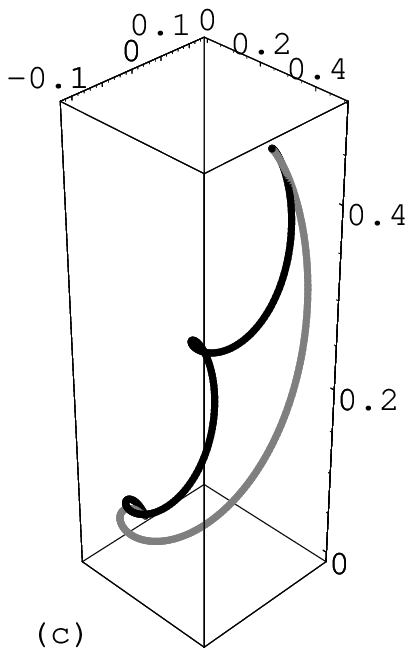}
  \includegraphics[height=95mm,width=50mm,clip]{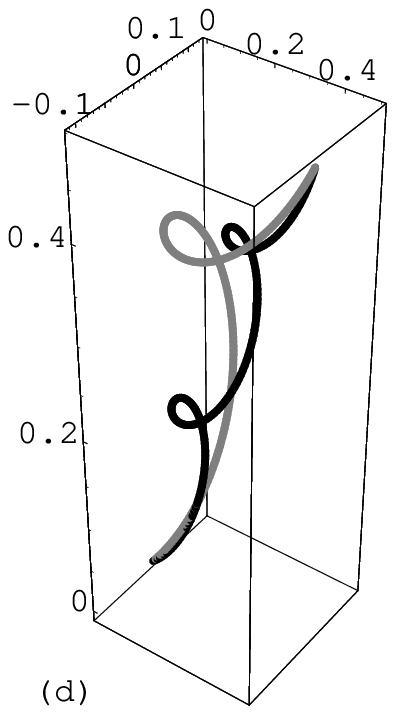}
  \end{center} \caption{Rods for $P_{\psi}=P_{\phi}$ and
  $F=1$. The lines corresponds to $P_{\phi}=P_{\psi}=10$ (black) and
  $P_{\phi}=P_{\psi}=5$ (gray). (a) and (b) show the up and down
  solutions for $r_f=0.39$ and $z_f=0.8$; (c) and (d)  show the up and
  down solutions for $r_f=0.5$ and $z_f=0.49$}
\label{fig7} 
\end{figure}

Figure \ref{fig6} shows the $(r_f,z_f)$ map for
$P_{\psi}=P_{\phi}=5$. It resembles the map on Fig. \ref{fig4}, only
distorted towards $z_f=1$. For these values of the parameters the rod
admits 'vertical' configurations, as opposed to the 'horizontal'
configurations displayed in the previous case. Figures \ref{fig7} show
examples of rods with $P_{\psi}=P_{\phi}=5$ and $P_{\psi}=P_{\phi}=10$
for $r_f=0.39$ and $r_f=0.5$.

\section{Application to Non-homogeneous DNA}  
 
As a last application of our method we shall consider the equilibrium
configurations of {\it non-homogeneous} rods. We restrict ourselves to
the simplest case of periodic non-homogeneities in Young's
modulus. The motivation for this study is the fact that repeated (and
therefore periodic) DNA sequences form a substantial fraction of all
eukaryotic genomes \cite{jenny,brian}. The calculations presented here
are based on the stiffness parameters recently computed for the 32
tri-nucleotide units from DNA data \cite{gromiha}. Our goal is to
understand how much the equilibrium configuration of a non-homogeneous
rod differs from that of the homogeneous case when the rod is subject
to fixed mechanical conditions.

Repetitive DNA is formed by nucleotide sequences of varying lengths
and compositions. Repeated sequences, reaching up
to 100 megabasepairs of length \cite{brian}, appear to have little
or no functional role, and are commonly regarded as ``selfish'' or
``junk'' DNA \cite{mclister}. We shall use a simple periodic formula
for the (scaled) Young's modulus that covers most of the parameter
interval spanned by the tri-nucleotides given in ref. \cite{gromiha}:
\begin{equation}  
\label{young}  
E(s)=1+\alpha \cos \frac{2\pi}{{\cal L}}s \;.
\end{equation}  
${\cal L}$ is the period of the oscillations of the Young's modulus
along the DNA and $\alpha $ is a parameter that depends on the
specific sequence being repeated and that can not be greater than
$0.66$.  
  
Eqs.(\ref{hamilton})-(\ref{Ppsi}) have to be slightly modified to
include the non-constant Young's and shear moduli. We obtain
\begin{equation}  
\label{hamilton2}  
H=\frac{P^{2}_{\theta}}{2E(s)}+\frac{P^{2}_{\phi}}{2\Gamma _{0}\mu (s)}+  
\frac{(P_{\psi}-P_{\phi}\cos \theta)^{2}}{2E(s)\sin ^{2} \theta}+  
F\cos \theta,  
\end{equation}
with  
\begin{equation}  
\label{Pteta2}  
P_{\theta}=E(s)\theta ',  
\end{equation}  
\begin{equation}  
\label{Pphi2}  
P_{\phi} \equiv M_{3}=\Gamma _{0}\mu (s)(\phi '+\psi '\cos \theta),  
\end{equation}  
\begin{equation}  
\label{Ppsi2}  
P_{\psi} \equiv M_{Z}=E(s)\psi '\sin ^{2}\theta + P_{\phi}\cos \theta .  
\end{equation}  
The solutions do not depend on $\mu(s)$, since it does not enter in
the differential equations for $\theta$, $P_{\theta}$ and
$\psi$. Notice that these equations are not integrable if $\alpha \neq
0$. Although $P_{\psi}$ and $P_{\phi}$ are still constants, the
elastic energy per unit arclength is not.

The method for solving the BVP for a non-homogeneous rod is the
following: consider a solution extending from the origin to ${\bf
r}_f$ with $\alpha=0$ and initial conditions $\theta_0$ and 
$P_{\theta_0}$. Now integrate the Kirchhoff equations
from the same initial conditions but using Eq.(\ref{young}) with
$\alpha \neq 0$. This new solution, whose end point is ${\bf
r}_f+\delta {\bf r}$, can be used as a trial solution for the
non-homogeneous rod. Using the method of section IV we push the rod
back to ${\bf r}_f$.

\begin{figure}[ht] 
  \begin{center}
  \includegraphics[height=95mm,width=50mm,clip]{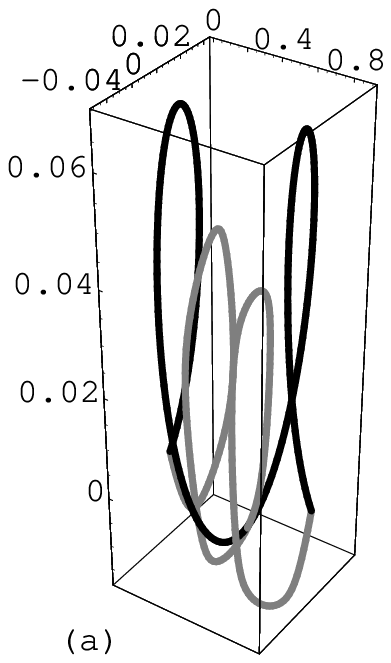}
  \includegraphics[height=95mm,width=50mm,clip]{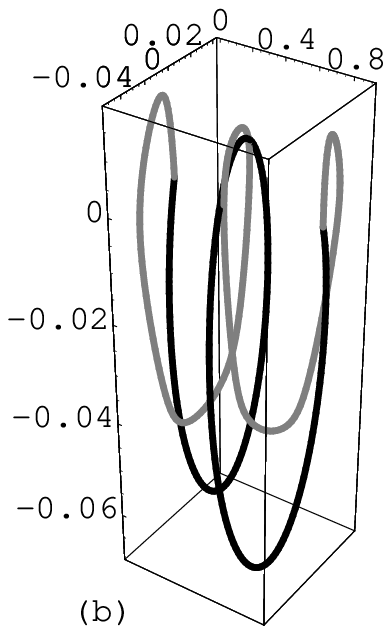}
  \end{center} \caption{Comparison between homogeneous and
   non-homogeneous rods. Panels (a) and (b) show the up and down 
   solutions respectively. The parameters are $P_{\psi}=0$,
   $P_{\phi}=10$, $F=1$, $r_f=0.9$ and $z_f=0$. The black curve shows
   the homogeneous rod and the gray curve the non-homogeneous rod with
   $\alpha=0.66$ and ${\cal L}=0.1$. }
\label{fig8} 
\end{figure}

Figure \ref{fig8} shows the up and down solutions for load parameters
$F=1$, $P_{\psi}=0$, $P_{\phi}=10$, $r_f=0.9$ and $z_f=0$ for the
homogeneous rod (black curve) and a non-homogeneous rod with
$\alpha=0.66$ and ${\cal L}=0.1$ (gray curve).

\begin{figure}[ht] 
  \begin{center}
  \includegraphics[height=95mm,width=50mm,clip]{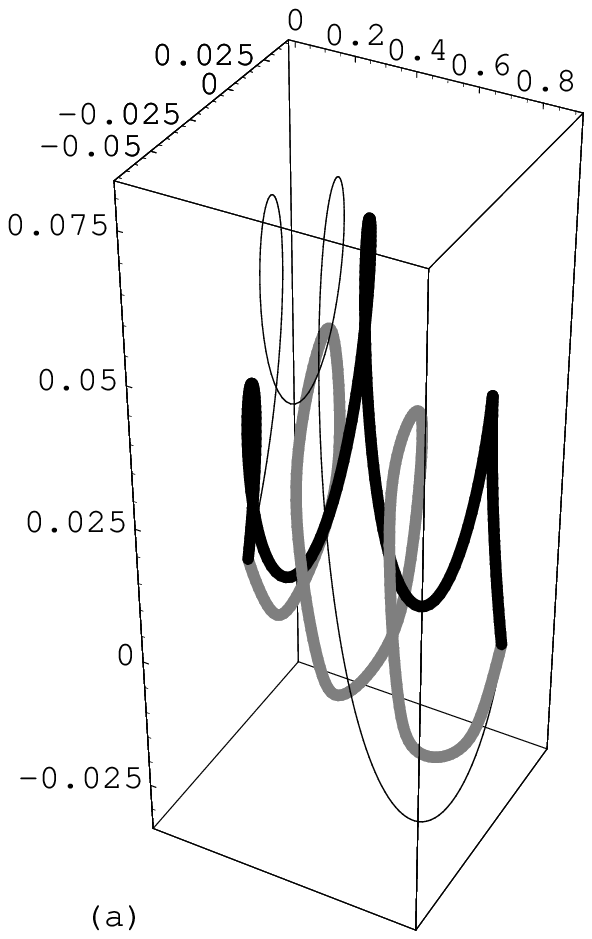}
  \includegraphics[height=95mm,width=50mm,clip]{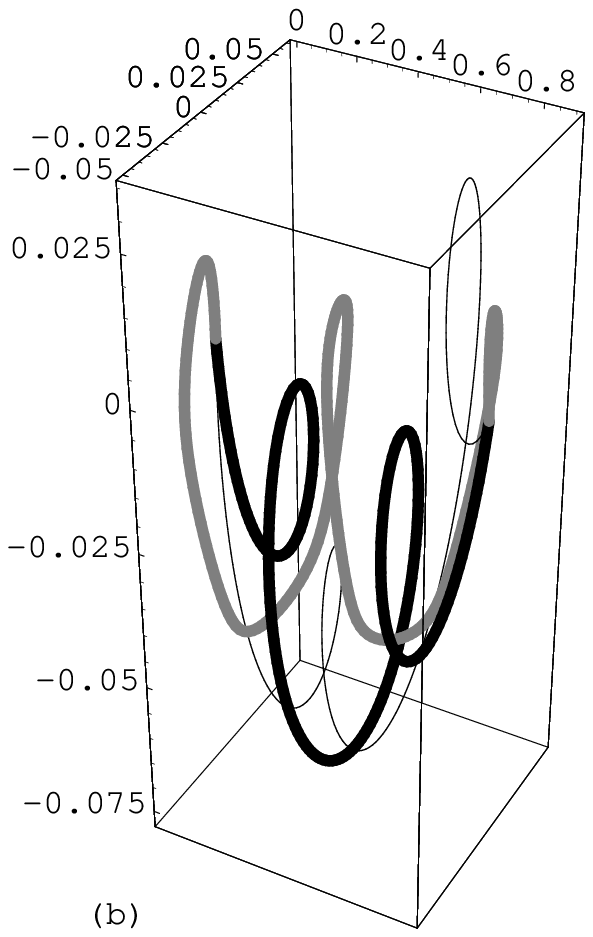}
  \end{center} \caption{Comparison between non-homogeneous rods with 
   different periods ${\cal L}$. Panels (a) and (b) show the up and down 
   solutions respectively. The load parameters and final position 
   $\mathbf{r}_f$ are the same as in Fig. \ref{fig8}. The curves show
   rods for ${\cal L}=0.1$ (gray), ${\cal L}=0.5$ (thick black) and
   ${\cal L}=0.65$  (thin black) }
\label{fig9} 
\end{figure}

Finally, Figure \ref{fig9} shows the effect of changing the period of
oscillations in Young's modulus. We show the up and down solutions for
non-homogeneous rods with the same load parameters of the previous
figure and $\alpha=0.66$. The curves show rods for ${\cal L}=0.1$
(gray), ${\cal L}=0.5$ (thick black) and ${\cal L}=0.65$ (thin
black). The changes in the three-dimensional shape of the rods are
evident for these values of load parameters. The sensitivity of the
three-dimensional shape to the base-pair sequences indicated that DNA
repeats may have conformational roles.

\section{Conclusions}  

In this work we presented a general method to solve the boundary value
problem (BVP) for finite Kirchhoff filaments. The method consists in
making small changes to a known {\it trial} rod that satisfies the
Kirchhoff equations but not necessarily the boundary conditions. We
combine a shooting technique with the method of monodromy matrix to
push the end point of the trial solution to the desired position, step
by step.  By linearizing the Kirchhoff equations we obtain an explicit
relation between a variation of the initial conditions (expressed in
terms of Euler angles) and the consequent variation of the rod's end
point.
  
The solutions of the BVP are limited by the physical constraints of
the rod, such as the moments and tension. A sketch of the allowed end
points can be constructed by integrating the Kirchhoff equations for a
large number of initial conditions. The regions of possible end points
form complex figures reflecting the non-linear character of the
equations. The regions of existence of end points may serve as a guide
to find the appropriate load parameters needed for a desired solution.

We presented several examples of rods with different end positions at
different distances from the origin for various sets of load
parameters. In all cases the method worked very well and the BVP was
solved with at least six significant figures in the values of $r_f$
and $z_f$. We also applied the method to non-homogeneous,
sequence-dependent, DNAs. We modeled pieces of repeated sequences by a
sinusoidal oscillation of the Young's modulus. We showed that
the tri-dimensional structure of the DNA is indeed sensitive to the
presence of such sequences, a property that has been considered before
\cite{hogan} but studied only in terms of the DNA intrinsic curvature
\cite{manning}. The effect studied here may contribute to other 
sequence-dependent properties that affect the three-dimensional
conformations of the DNA. 

\acknowledgments This work was partially supported by the Brazilian
agencies FAPESP, CNPq and FINEP.

\newpage  
 
\end{document}